\documentclass[aps,preprint]{revtex4-1}
\usepackage{ifpdf}

\usepackage[margin=1.0in]{geometry}
\usepackage[pdftex]{graphicx}
\usepackage{epstopdf}
\begin{document}

\title{Three-dimensional quantum size effects in Au islands on MoS$_2$}

\author{T.~E.~Kidd, J.~Weber, R.~Holzapfel, K.~Doore, and A.~J.~Stollenwerk}\email{andrew.stollenwerk@uni.edu}
\affiliation{University of Northern Iowa, Department of Physics, 215 Begeman Hall, Cedar Falls, Iowa 50614-0150, USA}

\date{\today}

\begin{abstract}

Quantum confinement was found to play a critical role in the formation of Au(111) islands grown on the surface of MoS$_2$. These confinement effects are fully three dimensional, with a strong correlation to the relatively large Fermi wavelength associated with the Au(111) planes. The confinement effects result in preferred heights with a periodicity of nearly 2~nm and persist to much higher temperatures than are typically seen in electronic growth mode systems. These findings indicate the potential to explore electronic growth modes in a new class of systems based on metal - layered semiconductor interfaces.

\end{abstract}



\maketitle

\section{Introduction}

Quantum well states can play a role in determining the geometric structure of nanometer scale metallic films.  Film stability increases as quantum well states shift farther from the Fermi level.  Energy contributions due to quantum confinement perpendicular to the substrate minimized when the film thickness is a half integer multiple of the Fermi wavelength, resulting in the opening of a gap at the Fermi level of the metal.  Such contributions to the electronic configuration cause oscillations in the total energy of the film with thickness.  The set of preferred or "magic" heights resulting from these energy considerations have been reported in various systems such as Pb/Si(111)~\cite{yeh:prl:85:5158, su:prl:86:5116, otero:prb:66:115401}, Ag/Si(111)~\cite{goswami:ss:601:603, gavioli:prl:82:129, han:prb:81:115462,  huang:ss:416:L1101, miyazaki:ss:602:276}, and Ag/Fe(100)~\cite{man:prb:81:045426,paggel:sc:283:1709}.

Theoretically, quantum size effects could be used to create atomically flat films with a high degree of uniformity, ideal for the creation of metallic contacts for either scientific investigations or device applications on the nanometer scale.  In practice, this can be challenging as the energies associated with quantum confinement are typically too small to make a significant impact, even at room temperature.  To observe quantum size effects the relatively large contributions from sources such as surface free energy, strain due to lattice mismatch, and surface kinetics must be minimized.  This has been accomplished through cryogenic deposition using materials with either a high degree of lattice matching~\cite{man:prb:81:045426} or materials that form a wetting layer~\cite{miyazaki:ss:602:276} that maintain confinement and reduce strain at the interface.  This imposes a serious constraint on the number of acceptable systems when taking into account the need for strong confinement.  The surface of layered semiconductor crystals have many properties favorable for electronic confinement.  In most cases, the interface between metals and these materials is separated by a van der Waals gap~\cite{allain:nm:14:1195}, resulting in an abrupt interface with minimal strain due to lattice mismatch.  This provides the possibility of achieving quantum size effects without the afore mentioned constraints.

In this paper, we present scanning tunneling microscopy (STM) evidence that quantum size effects play a defining role in the morphology of Au(111) islands on MoS$_2$.  Room temperature deposition results in atomically smooth triangular structures with a Gaussian height distribution.  Upon annealing, these structures coalesce into islands with quantized dimensions that correlate strongly with a portion of the Au Fermi surface.

\section{Experimental}

MoS$_2$ substrates were purchased from commercially available sources similar to those used in other recent studies~\cite{radisavljevic:nn:6:147, mcdonnell:nano:8:2880, cook:prb:92:201302}.  The surface was cleaned via mechanical exfoliation prior to being introduced into the load lock of a home-built thermal deposition chamber with base pressure of 2~$\times$~10$^{-9}$~mbar.  Each MoS$_2$ substrate was degassed for one hour at 500~K as measured by a type K thermocouple attached to the sample holder.  Approximately 3.0~nm of Au was deposited at rate of 0.01~nm/s measured $in$ $situ$ using a quartz microbalance.  After deposition, samples were attached to STM appropriate sample holders $ex$ $situ$ using a conductive silver paste and introduced into a variable temperature UHV STM system (Omicron) with base pressure of 7$\times$10$^{-9}$~mbar.  Tips were made from a W wire electrochemically etched in a 5~M potassium hydroxide solution with a 5~V$_{rms}$ bias.  Differential spectra were extracted at room temperature using a lock-in amplifier with a 3.5~kHz, 76~mV modulation.

Beginning at ambient temperature, the sample was heated in steps of approximately 25~K as measured by a Pt100 thermistor in contact with the imaging stage.  The surface was scanned at each incremental temperature up to 575~K, the maximum temperature of the imaging stage.  Scanning was paused after each temperature increase to allow the sample to achieve thermal equilibrium.  In ordered to reach higher temperatures, the sample was transferred to a separate preparation stage located adjacent to the STM head.  Samples were annealed at temperature for approximately five minutes and cooled to room temperature before being transferred back to the imaging stage to scan the surface.  This was done in 75~K increments up to 725~K, the highest effective working temperature of the conductive paste (i.e. electrical conduction is lost with the sample).

\section{Results and Discussion}

\begin{figure}[h]
\includegraphics{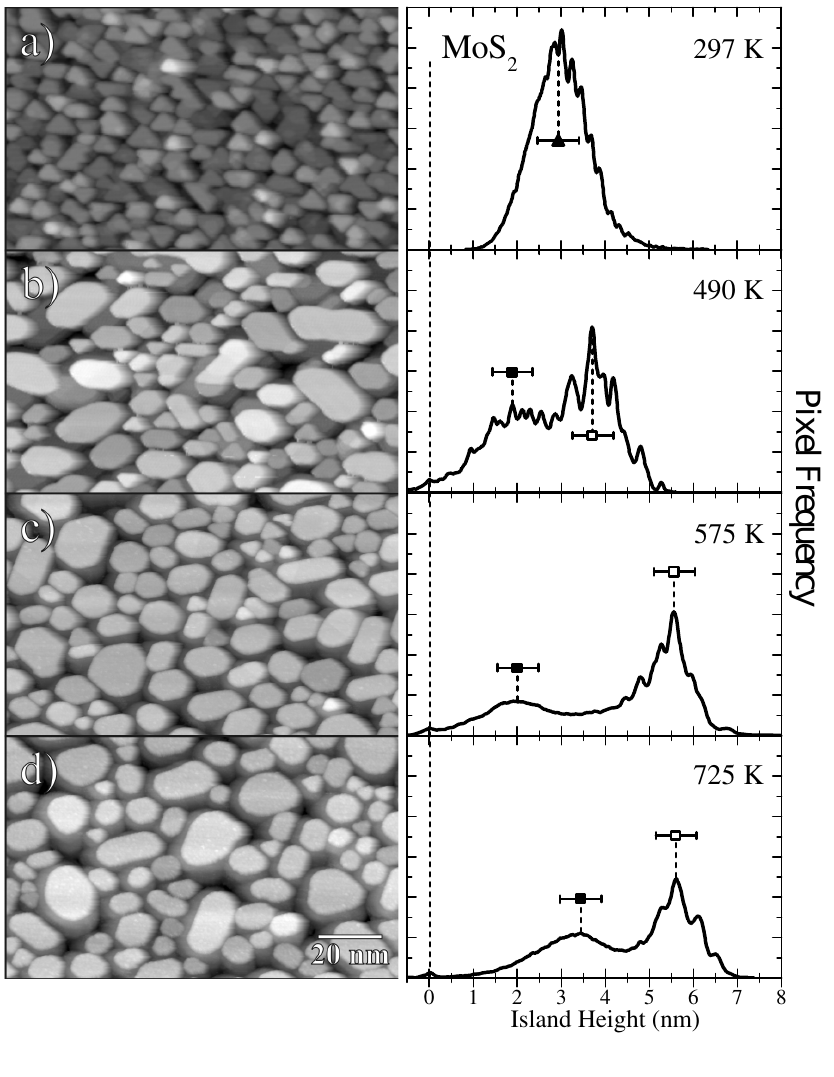}
\caption{STM topography of 3.0~nm Au on MoS$_2$ at room temperature (a) and after annealing (b)-(d). Hight distributions are depicted to the right of each image.  Preferred heights are labeled as closed squares for the lower peaks and open squares for the upper.  All images were acquired at V$_{tip}$ = 0.7~V and I$_{tip}$ = 0.9~nA.} \label{image}
\end{figure}

Prior to deposition, MoS$_2$ substrates were found to be extremely flat with several microns between step edges.  The image in Fig.~\ref{image}(a) is representative of the surface after room temperature deposition and consists primarily of atomically flat, terraced triangular structures with orientations reflecting the sixfold symmetry of the substrate.  The integer step height of these terraced structures corresponds with Au(111), in agreement with previous observations~\cite{darby:pssa:25:585, honjo:pssa:55:353, nagashima:jcg:146:266}.  Figs.~\ref{image}(b)-(d) show how the triangular structures coalesce into a combination of larger triangles and semi-hexagonal structures as the sample temperature increases to 725~K.  Occasionally, entire islands spontaneously disappeared from the surface under normal scanning conditions, demonstrating the weak bonding at the interface.

\begin{figure}[h]
\includegraphics{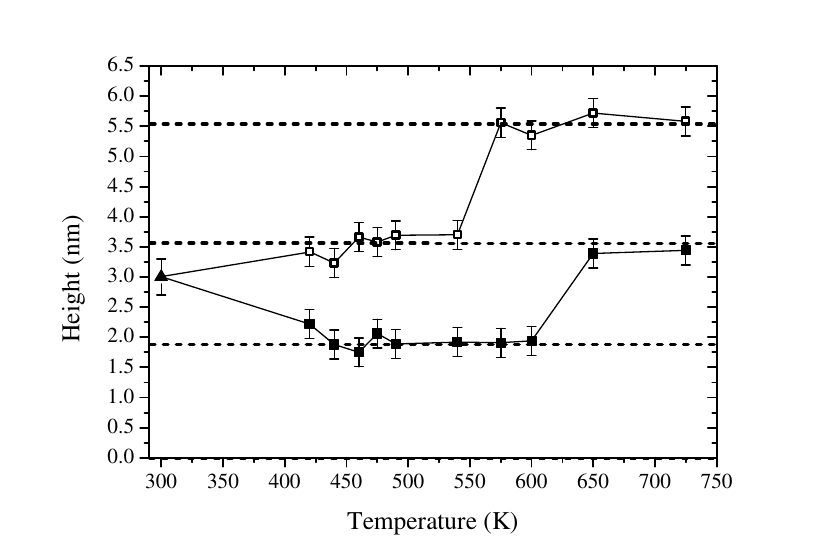}
\caption{Preferred island heights as a function of temperature.} \label{peaks}
\end{figure}

The height distributions calculated from these images are depicted on the right side of Fig.~\ref{image} and are comparable to those calculated from other images.  The film has a highly symmetric distribution of heights immediately after deposition as seen in Fig.~\ref{image}(a).  Due to the high degree of symmetry, it was assumed that this distribution was centered on the nominal deposition thickness of 3.0~nm.  As the sample is annealed to 400~K, the Au film begins to acquire an asymmetric bimodal distribution with a preference for greater heights.  As the islands grow in height, the substrate is exposed and produces an MoS$_2$ peak that is first visible at 420~K.  The MoS$_2$ peak is used as a reference to determine the absolute height of the Au islands for each annealed data set. When possible, the preferred heights are determined by the location of individual monolayer peaks found in the distribution.  When individual monolayers could not be distinguished, a Gaussian peak analysis was used.  The location of the first and second peaks are labeled in Figs.~\ref{image}(b)-(d) as closed and open squares, respectively.  Preferred heights are plotted as a function of temperature in Fig.~\ref{peaks}.  Similarly, the closed squares represent the lower peak and the open squares represent the higher peak.  Error in the measurement of these preferred heights is estimated to be no more than $\pm$~0.24~nm (a single atomic layer). From Fig.~\ref{peaks} three stable heights are apparent occurring at 1.90, 3.51, and 5.55~nm, giving an average interval of 1.85~$\pm$~0.24~nm.  Shifts between these heights with increasing temperature appear to happen relatively abruptly.

\begin{figure}[h]
\includegraphics{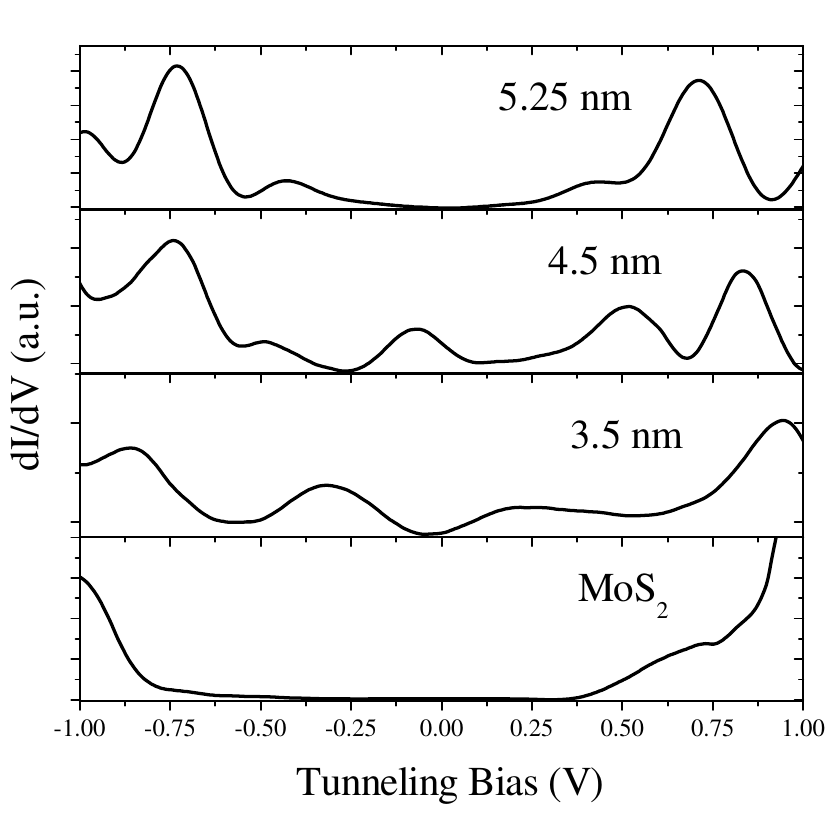}
\caption{Differential tunneling spectroscopy obtained on the MoS$_2$ surface and Au islands of different heights.  Tip stabilization parameters for each spectrum were V$_{tip}$ = 0.76~V and I$_{tip}$ = 2~nA. }\label{spectra}
\end{figure}

Differential tunneling spectroscopy obtained from various Au islands are shown in Fig.~\ref{spectra} in addition to a reference spectrum from the MoS$_2$ surface.  The reference spectrum is similar to those obtained previously on bulk MoS$_2$~\cite{fuhr:prl:92:026802}.  Spectra from the bulk Au(111) surface are relatively featureless in this energy range with the exception of a prominent surface state approximately 0.5~eV below the Fermi level~\cite{kim:prb:80:245402}.  Unlike the bulk, the spectra from the Au islands exhibit a significant amount of structure consistent with resonant states observed in quantum wells or dots.  Spectra taken on islands with preferred heights of 3.5 and 5.25~nm show minimal density of states at the Fermi level. A large density of states at the Fermi level can be seen in spectra taken from a 4.5 nm high island, a height that does not commonly occur.  These results show that the preferred heights represent minima in the electronic energy of the system.

The 1.85~nm periodicity is far too large to be the result of the isotropic Au Fermi wavelength ($\lambda_F/2 \approx$ 0.25~nm).  Within the \{111\} planes $\lambda_F/2$ is, however, approximately 1.45~nm for the bulk~\cite{coleridge:prb:25:7818} and 1.75~nm for the surface (averaging the splitting of the twinned surface state with $\lambda_F/2 \approx$ 1.85 and 1.65~nm)~\cite{reinert:prb:63:115415}.  The surface wavelength is nearly a perfect match to the data.  It is worth noting that the Au islands of this system are not truly representative of a bulk material. In fact, the value of the bulk $\lambda_F$ is found to be nearly equal to that of the surface in Au(111) films as thick as 140~nm~\cite{schouteden:prb:79:195409}.

\begin{figure}[h]
\includegraphics{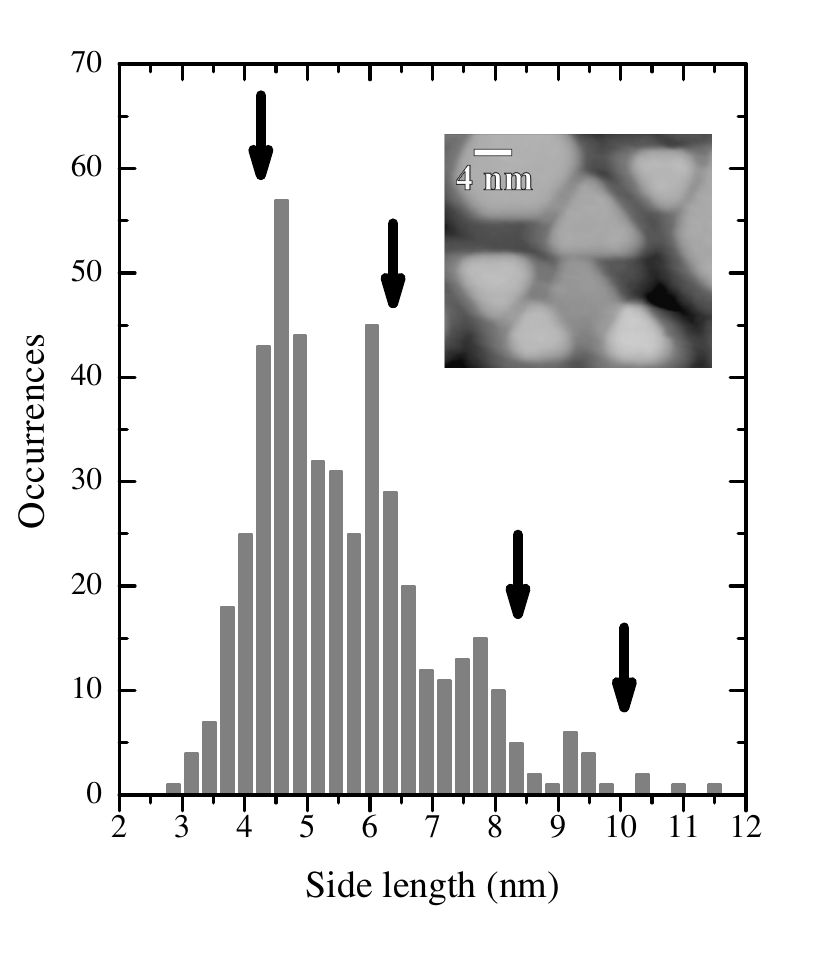}
\caption{ The distribution of triangular island side lengths.  Side lengths resulting in quantum well states farthest from the Fermi level are labeled by vertical arrows (see manuscript for description of model).   }\label{Lateral}
\end{figure}

The lateral sizes of these islands also exhibited quantization correlated with $\lambda_F$ of the Au(111) surface. The distribution of side lengths for the equilateral triangular islands is plotted in Fig.~\ref{Lateral} and shows four preferred lengths up to 12~nm. Knowing that the preferred physical dimensions occur when quantum well states are furthest from the Fermi level, a model was developed to compare with these results.  In this model, the islands are approximated as 2D potential wells in the shape of an equilateral triangle.  Using the Au(111) surface state, the energy parallel to the (111) plane is expressed in terms of the effective mass ($m^*$ = 0.255$m_0$) and the band minimum ($U_0$ = 0.487~eV)~\cite{reinert:prb:63:115415} as,
\begin{equation}
    E(k_{||}) = -U_0 + \frac{\hbar^2k_{||}^2}{2m^*}.
\end{equation}
The depth of the potential well was determined by the sum of the (111) band minimum and the work function of Au taken to be W = 5.31~eV~\cite{lyon:apl:88:2208267}.  The side length of the triangular wells was set to integer multiples of the Au(111) lattice constant (0.288~nm), up to 12~nm.  Each well was surrounded by a vacuum layer no less than 25\% of the side length to account for wave penetration into the forbidden region.  The Schr\"{o}dinger equation was solved for each potential well using an eighth order finite difference method with a 750$\times$750 mesh. Convergence of our results with respect to the mesh size was thoroughly tested.  Side lengths resulting in quantum well states farthest from the Fermi level are labeled by vertical arrows in Fig.~\ref{Lateral}. The model produces the expected four preferred side lengths and corresponds quite well to the strongest features at 4.5 and 6.0~nm.

The Au islands exhibit both lateral and vertical quantization correlated with the Au(111) Fermi surface.  Lateral quantization naturally couples to these states, as has been seen in Bi nanostructures with quantized widths grown on HOPG~\cite{kowalczyk:nl:13:43}. It is less clear how the Au(111) states could result in vertical quantization of the island heights. In fact, owing to the necks in the Au Fermi surface in the (111) plane, there are no states directed purely in the surface normal direction. However, the (\={1}11), (1\={1}1), and (11\={1}) planes are oriented at a 19 degree angle from the surface normal.  The vertical component of these states would give a Fermi wavelength somewhat larger than the bulk \{111\} planes that would provide the necessary criteria for electronic quantization of the island heights.

\section{Conclusions}

In summary, we have observed that the growth of Au(111) islands on MoS$_2$ exhibits quantum size effects in three-dimensions.  The island heights and lateral dimensions correlate well with the Fermi wavelength in the Au\{111\} planes, indicating that quantization is driven by quantum well states.  These effects persist to at least 725~K, an unusually high temperature for an electronic growth mode.  This is likely due to the weak bonding at the Au/MoS$_2$ van der Waals interface, which minimizes strain but can still induce epitaxial growth without a wetting layer or significant lattice matching.  The abrupt van der Waals gap at the interface also gives rise to strong electronic confinement, ideal for the formation of quantum well states.  While further study is needed to fully understanding of the microscopic origins of this quantization, this discovery creates a new class of magic size systems and a new route for the control of metal film growth parameters on van der Waals surfaces.

\section{Acknowledgments}
This work was supported by the National Science Foundation Grant No. DMR-1410496.
 \bibliographystyle{achemso.bst}
\pagebreak

\pagebreak

\centerline{\textbf{FIGURE CAPTIONS}}

\end{document}